\documentclass[aps,prl,twocolumn, showpacs,superscriptaddress]{revtex4-1}

\usepackage{amsfonts}
\usepackage{amsmath}
\usepackage{amssymb}
\usepackage{graphicx}%
\usepackage{soul}
\usepackage{color}\pagecolor{white}
\usepackage{ulem}
\newcommand{\bra}[1]{\ensuremath{\left\langle#1\right|}}
\newcommand{\ket}[1]{\ensuremath{\left|#1\right\rangle}}

\begin{document}

\title{Testing Time Reversal Symmetry in Artificial Atoms}

\author{Frederico Brito}
\email[]{fbb@ifsc.usp.br}
\affiliation{Instituto de F\'{i}sica de S\~ao Carlos, Universidade de S\~ao Paulo, C.P. 369, 13560-970, S\~ao Carlos, SP, Brasil}

\author{Francisco Rouxinol}
\affiliation{Department of Physics, Syracuse University, Syracuse New York 13244-1130, USA}

\author{M. D. LaHaye}
\affiliation{Department of Physics, Syracuse University, Syracuse New York 13244-1130, USA}

\author{Amir O. Caldeira}
\affiliation{Instituto de F{\'\i}sica Gleb Wataghin, Universidade Estadual de Campinas-UNICAMP, 13083-859 Campinas, SP, Brazil}


\begin{abstract}
Over the past several decades, a rich series of experiments has repeatedly verified the quantum nature of superconducting devices, leading some of these systems to be regarded as artificial atoms. In addition to their application in quantum information processing, these `atoms' provide a test bed for studying quantum mechanics in macroscopic limits. Regarding the last point, we present here a feasible protocol for directly testing time reversal symmetry through the verification of the microreversibility principle in a superconducting artificial atom. Time reversal symmetry is a fundamental property of quantum mechanics and is expected to hold if the dynamics of the artificial atom strictly follow the Schr\"odinger equation. However, this property has yet to be tested in any macroscopic quantum system. In the end, as an application of this work, we outline how the successful implementation of the protocol would provide the first verification of the quantum work fluctuation theorems with superconducting systems.
\end{abstract}

\maketitle

\section{Introduction}

Few concepts in nature are so simple and yet as profound as those related to symmetry. Indeed, the beauty of its manifestations has led to the modern view that principles of symmetry dictate the forms of nature's fundamental laws\cite{Gross}, embodying striking implications that range from conservation principles to the classification of elementary particles.

Time reversal symmetry (TRS) is a prominent example that underlies a large variety of phenomena. In many instances, the fundamental microscopic laws of nature are invariant under time reversal transformations. This invariance is at the heart of microscopic reversibility (microreversibility)\cite{Messiah}, which itself is crucial to powerful concepts such as the principle of detailed balance\cite{Reichl}, the fluctuation-dissipation theorem\cite{Callen}, and  fluctuation relations (e.g. Tasaki-Crooks fluctuation theorem)\cite{Campisi}, to name a few.

Yet TRS is not an exact symmetry of nature: in the very least, it is observed to be broken in elementary processes that involve the weak interaction\cite{BABAR,Bernabeu}, and moreover, there is evidence to suggest that it must also be violated over a much broader range of conditions in order to account for the prevalence of matter over anti-matter in the universe\cite{Sakharov,henley}. Manifestations of such violations potentially herald new phenomena and are thus the subject of extensive experimental investigations in both atomic and particle physics\cite{henley}.  

While considerable effort has been invested in the search for violations of TRS in the interactions of fundamental particles, experiments have not been conducted to investigate TRS in the physics of quantum systems at the macroscopic scale. Specifically, the question thus arises: Once one properly takes into account dissipative and decoherence effects, would TRS be observed in, say, a mesoscopic or even a macroscopic device? On the face of it, there is no reason to expect that it breaks down: we know the microscopic laws of quantum mechanics can be applied to at least some (properly prepared) macroscopic systems. Nonetheless, if it does break down, this must reflect new physics, which could have potential connections to open questions like the nature of the quantum-classical divide\cite{LeggettJP,Ball}.

With these thoughts in mind, we delineate here a protocol for directly testing TRS in an artificial atom that is based upon a superconducting quantum device (SQD). While similar types of SQDs are known for their use as qubits in the development of quantum computing architectures\cite{Clarke,Devoret}, we propose to utilize an SQD as a multi-level artificial atom to test a specific manifestation of TRS, namely the principle of microreversibility.

\section{Microreversibility and the artificial atom}
Generally speaking, the principle of microreversibility states that for each process (or trajectory in state space) that is accessible to a given system, there is an equally probable time-reversed process that the system can undergo\cite{Messiah}. In the context of quantum mechanics, it manifests in a simple relationship for the transition probabilities between any two states of a system whose Hamiltonian has undergone a time-dependent transformation\cite{Campisi}, namely that
\begin{equation}
P_{m|n}[\lambda]=P_{n|m}[\tilde{\lambda}]
\label{MR}
\end{equation}
where $P_{m|n} (P_{n|m})$ is the probability for the system to make a transition to state $\ket{m} (\ket{n})$ when it starts in state $\ket{n} (\ket{m})$. Here $\lambda$ represents the forward-in-time transformation of the system's Hamiltonian and $\tilde{\lambda}$ represents the motion-reversed process (Fig. 1a). It is important to note that the standard presentations of TRS (and consequently microreversibility) are done in the context of nondriven conservative systems \cite{Messiah}. However, as shown in Appendix A, the microreversibility principle can be readily adapted to include driven Hamiltonians, where the key element for recovering the standard relations consists in the temporal inversion of the Hamiltonian's temporal sequence\cite{Campisi}.

Equation \ref{MR} is a fundamental and general result for non-dissipative quantum mechanical systems, deriving from the invariance of a system's Hamiltonian under transformations by the anti-unitary time-reversal operator $\Theta$ \cite{Messiah,Campisi}. Thus it should hold true for all quantum systems in which TRS is maintained. Naively, one would expect this to include macroscopic systems for which the laws of quantum mechanics have been shown to apply, such as mechanical quantum systems\cite{Cleland,Palomaki} and superconducting cavities, circuits and devices\cite{Haroche, Clarke, Mooij,Houck,Devoret}. However, a direct test of TRS in these systems has yet to be performed. 

Concerning the role played by the macroscopic nature of the system, it is worthy of mentioning that the test of TRS we envision here has a different perspective than those conducted in other condensed matter systems. In fact, while here we want to address the emergence of TRS in quantum systems whose dynamics necessarily have to be described by the superposition of macroscopically distinct states or, at least, by a collective variable which obeys quantum mechanical laws, other studies have utilized macroscopic systems in order to magnify possible effects due to microscopic time-reversal violations\cite{Leggett}. One example is the search for permanent electric dipole moment (EDM) of elementary particles\cite{Ramsey} through measurements of the bulk magnetization of a macroscopic collection of spins\cite{Budker, Eckel}. Such experiments exploit the macroscopic size of the sample to significantly improve the signal acquisition, which is used to set limits on the existence of such permanent EDMs, allowing one to draw conclusions about the fundamental time-reversal invariance of the constituent elementary particles; by contrast, in our proposal, we conceive testing time-reversal invariance in the dynamics of a macroscopic degree of freedom representing the collective behavior of the constituent particles. 

As we show now, it should be technologically feasible to perform a test of microreversibility, and hence TRS, via Eq. \ref{MR}, in an artificial atom based upon an SQD, whose quantum dynamics is associated with circuit excitations, characterized by superpositions of several charge states.

The SQD here is a Cooper-Pair box (CPB), which in our proposal consists of a nanofabricated superconducting island (or box) that is formed by a pair of Josephson junctions in a DC SQUID configuration (Fig. 1b). The system is well-characterized by the following Hamiltonian\cite{Makhlin}\begin{figure}[t]
\begin{center}\includegraphics[ width=.8\columnwidth,keepaspectratio]{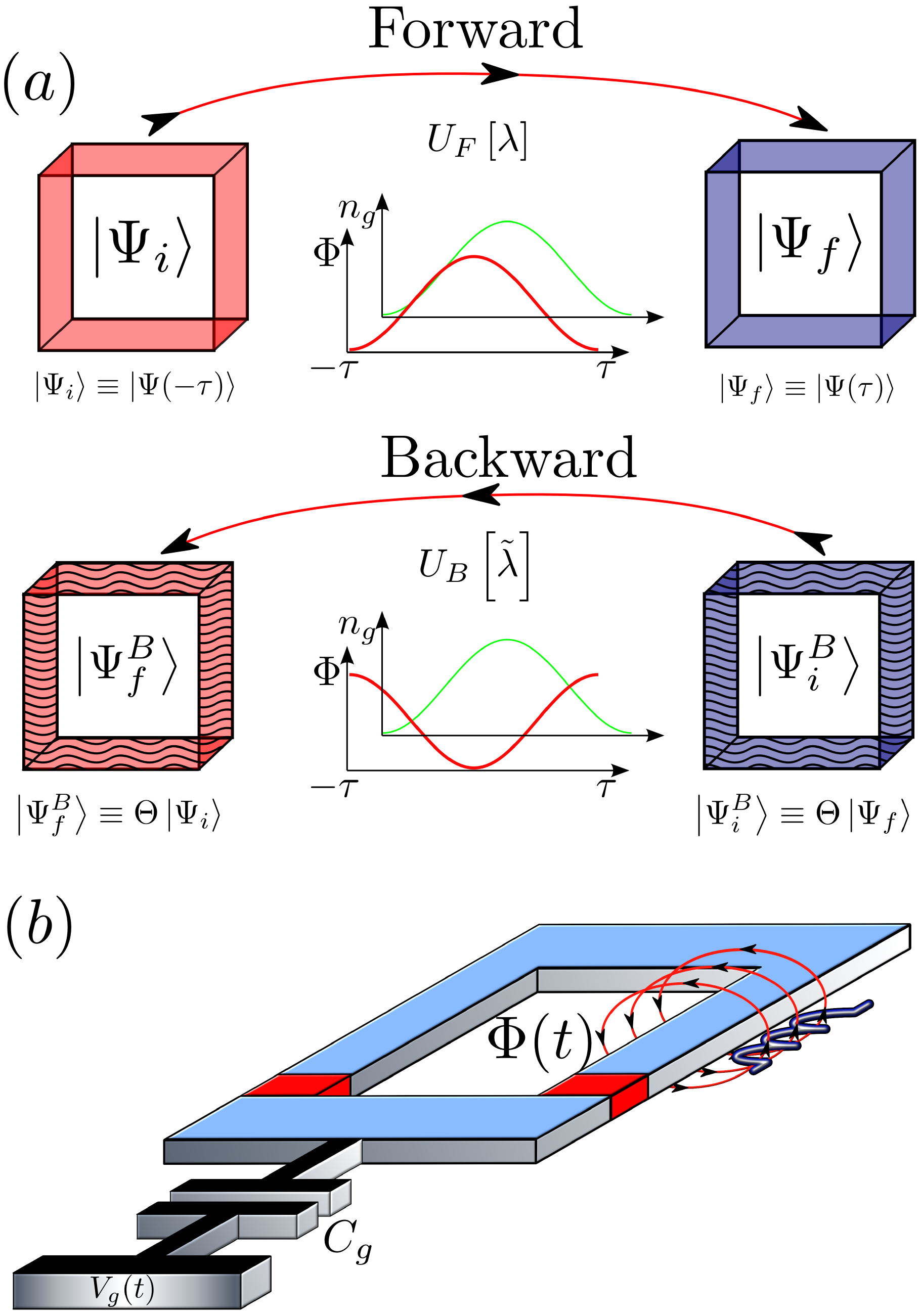}\end{center}
    \caption{Schematics of TRS for driven (nonautonomous) quantum systems and the Artificial Atom. (a) While the unitary time evolution of the forward-in-time protocol takes the initial state $\ket{\Psi_i}$ to the evolved state $\ket{\Psi_f}=U_F[\lambda]\ket{\Psi_i}$, the motion reversed state follows the dynamics $\ket{\Psi_f^{B}}\equiv\Theta\ket{\Psi_i}=U_B[\tilde{\lambda}]\Theta\ket{\Psi_f}$, where $\Theta$ represents the time reversal operator. In this generalization, $\Phi(t)$ and $n_g (t)$ represent time-dependent parameters in the system's Hamiltonian which are tuned to change the state of the system. If a system parameter depends upon an applied magnetic field, then the field must be inverted to move from the Forward to the Backward protocol as shown schematically with $\Phi(t)$. (b) SQD based upon a CPB, used to implement the artificial atom in our protocol. The system dynamics can be controlled by adjusting the magnetic flux $\Phi(t)$ through the loop and the charge $n_g(t)=C_gV_g(t)/2e$ on a nearby electrode, where $C_g$ is the capacitance of the CPB to the electrode and $V_g(t)$ is an externally controlled voltage.The device features two Josephson junctions (red boxes), arranged in parallel, interrupting the loop. The physical dimensions assumed here are such that the geometrical inductance is negligible compared to the Josephson inductances, leading to the Hamiltonian Eq. (\ref{Hamiltonian}), where: $E_C/\hbar= 2\pi \times3 {\rm GHz},~ E_{J\Sigma}/\hbar=2\pi\times10 {\rm GHz}$, and $\alpha=0.05$.}
    \label{fig1}
\end{figure}\begin{multline}
H=4E_C\sum_n(n-n_g)^2\ket{n}\bra{n}\\
-\sum_n\left[\frac{{\cal E}_J(\Phi)}{2}\ket{n}\bra{n+1}+\frac{{\cal E}_J^\ast(\Phi)}{2}\ket{n+1}\bra{n}\right].
\label{Hamiltonian}
\end{multline}The first term on the right-hand side of Eq. \ref{Hamiltonian} represents the electrostatic energy of the CPB for a given charge state $n$ (a discrete index labelling the number of Cooper-pairs on the island) and continuous polarization charge $n_g$ on a nearby electrode; the pre-factor $E_C$ is the total charging energy of the CPB. The second term on the right in Eq. \ref{Hamiltonian} represents the mixing of charge states due to the Josephson coupling of each junction. Here ${\cal E}_J(\Phi)\equiv E_{J\Sigma}\left\{\cos\left (\pi\frac{\Phi}{\Phi_0}\right)+i\alpha\sin\left (\pi\frac{\Phi}{\Phi_0}\right)\right\}$ is the total Josephson energy of the two junctions; observe that $|{\cal E}_J(\Phi)|$ is periodic in applied magnetic flux $\Phi$ with a period of one flux quantum $\Phi_0$. To account for asymmetry between the junctions, we define the parameter $\alpha\equiv (E_{J1}-E_{J2})/E_{J\Sigma}$, where $E_{J\Sigma}\equiv E_{J1}+E_{J2}$ is the sum of the individual junction Josephson energies $E_{J1}$ and $E_{J2}$.

It is important to note that numerous experiments over the past 15 years have shown that the two parameters $\Phi$ and $n_g$ in Eq. \ref{Hamiltonian} can be tuned {\it in situ} for experimental implementation of unitary operations with the CPB \cite{Clarke,Devoret}. The proposal we put forth for testing Eq. \ref{MR} exploits this coherent control. Specifically, it relies upon the adjustment of $\Phi$ and $n_g$ to modify the characteristics of the CPB's energy eigenstates. To understand how this might work, observe that, when $\Phi$ is adjusted so that $|{\cal E}_J|$ is relatively small (i.e $\beta\equiv|{\cal E}_J |/(4E_C )\ll1$), and $n_g$ is adjusted near an integer, the eigenstates of the system are essentially the charge states $\ket{n}$. On the other hand, if $\beta\gtrsim1$, or $n_g$ is near a half-integer, then the eigenstates are no longer well-defined charge states, but instead are weighted superpositions of multiple values of $\ket{n}$. Thus, through the rapid tuning of $\Phi$ and $n_g$, the CPB can be forced to undergo unitary evolution between various superpositions of charge states. Through repeated projective measurements of the CPB's charge state before $(n)$ and after $(m)$ identical forcing protocols, the transition probabilities $P_{m|n}$ between any given pair of charge states $\ket{n}$ and $\ket{m}$ in the spectral decompositions of the initial and final states can be constructed.

At this point it should be stressed that the kind of quantum states we envision using in our protocol to test TRS are not strictly speaking macroscopic in the same sense as the so-called ``cat states" \cite{Blatter,CaldeiraBook}. Whereas the latter are also present in superconducting devices when one studies, say, macroscopic quantum coherence in flux qubits or engineers entanglement between a superconducting microwave cavity and transmon qubit\cite{Sun}, the quantum state of the CPB in our protocol may involve the superposition of only a few charge states. Nevertheless, this device is macroscopic in the sense that it is an engineered system consisting of billions of atoms; and it is thus remarkable that a single collective variable still describes the dynamics of the device through genuine superpositions of its eigenstates. Moreover, it is also worth mentioning that although these states are susceptible to the influence of external interactions, one can operate the system under conditions(see below) which strongly reduce it and, therefore, safely describe its dynamics as unitary.

\begin{figure}[t]
\begin{center}\includegraphics[ width=0.9\columnwidth,keepaspectratio]{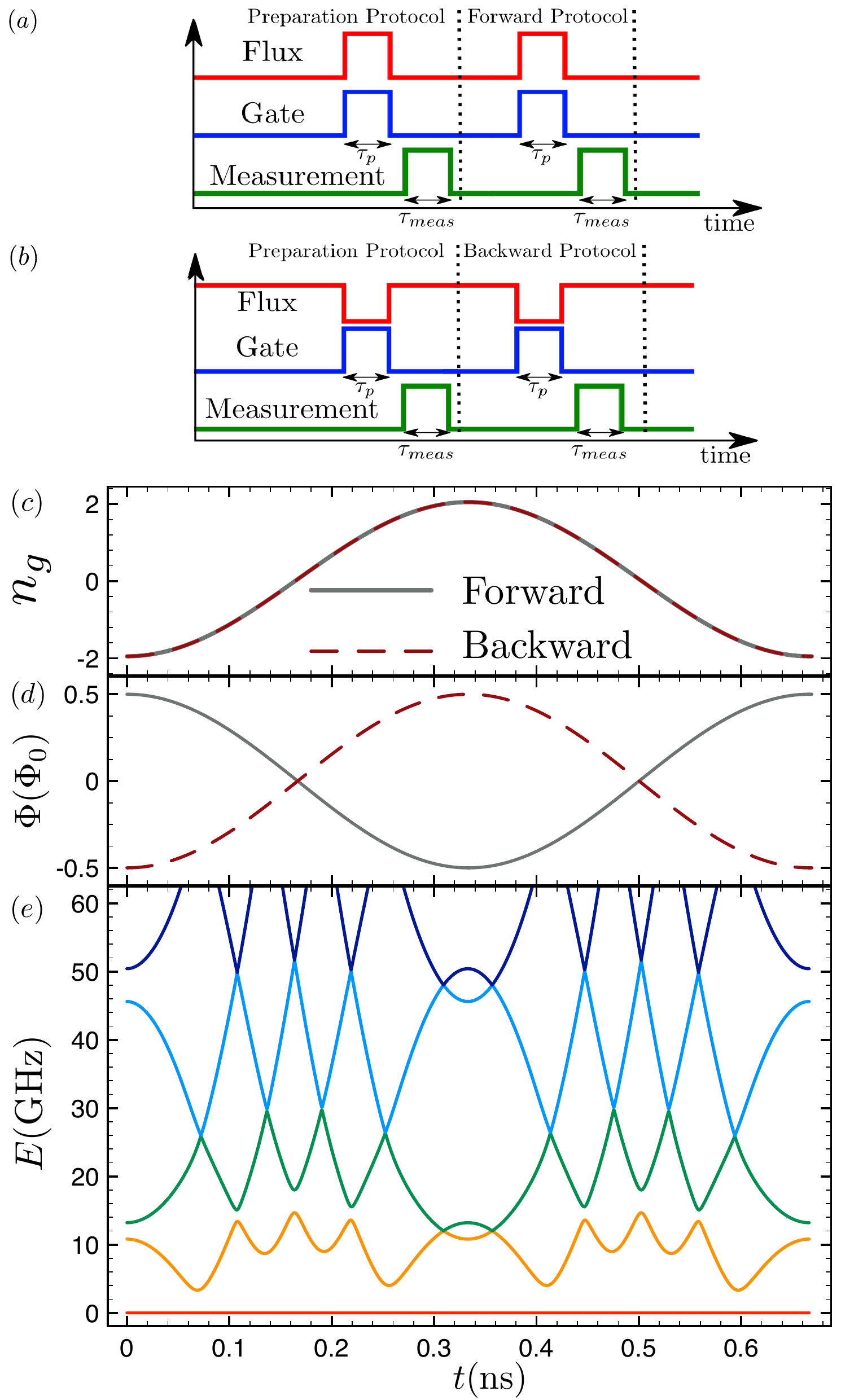}\end{center}
  \caption{Protocol scheme and system's eigenenergies versus time. (a-b), Outline of the Forward and Backward protocols. The first step, the Preparation Protocol, is used to construct an initial ensemble of several charge states. After the first measurement is performed, a driving protocol is used to implement a forward and backward-in-time evolution, which is followed by another charge measurement. After many runs, the transition probabilities between the allowed initial and final states can be constructed. (c-d), The time forward (solid line) and backward (dashed line) drive protocols for the gate charge $n_g$ and flux $\Phi$. In order to maintain the time reversal symmetry, the sign of the magnetic field must be inverted. (e), The eigenenergies of the CPB as function of time the driving protocol (the ground state energy is set zero). It is worth noticing the presence of several avoided level crossings, where Landau-Zener transitions are induced. The eigenenergies are calculated for the same parameters stated in Fig. 1.}
    \label{fig2}
\end{figure}

\begin{figure*}[]
\begin{center}\includegraphics[ width=2\columnwidth,keepaspectratio]{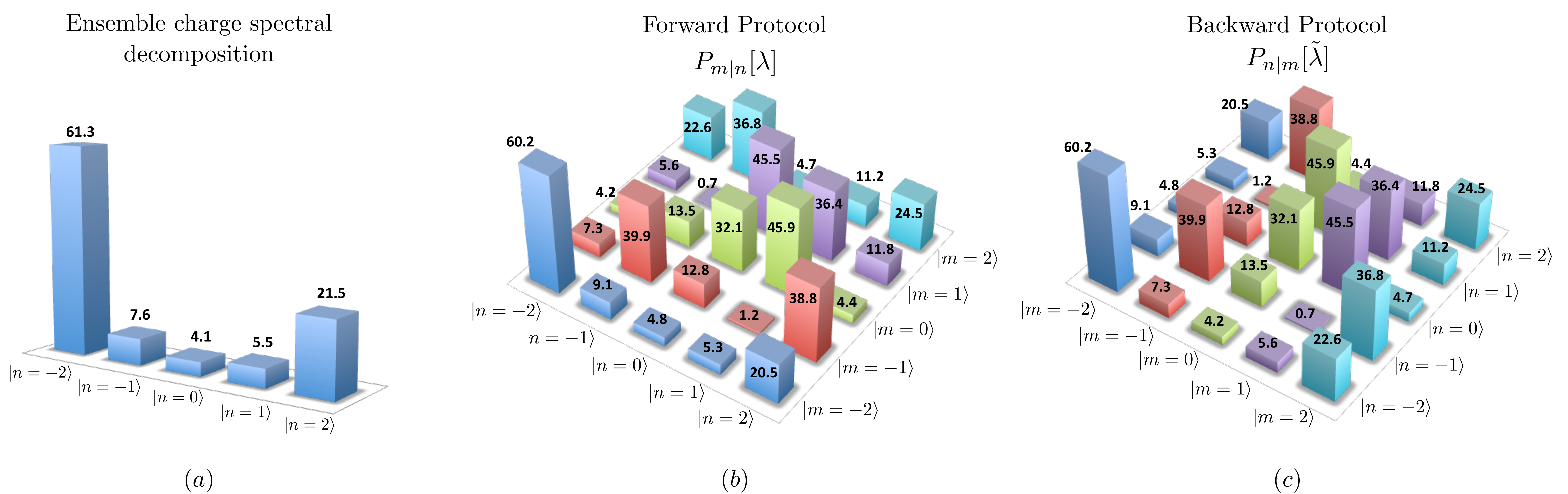}\end{center}
 \caption{Compiled probability distributions for preparation, forward and backward protocols. (a), The charge ensemble distribution $(\%)$ prepared after starting the system in the ground state and performing the Preparation Protocol. For the parameters considered here, the spectral decomposition obtained is predominantly $(> 99.9\%)$ comprised of charge states $\{-2,-1,0,1,2\}$. (b-c), The probability transitions $P_{m|n}$ $(\%)$ between the initial $\ket{n}$ and final $\ket{m}$ charge states determined for the Forward $\lambda$ (Backward $\tilde{\lambda}$) protocol. The leakage probability of leaving the charge subspace $\{-2,-1,0,1,2\}$ is determined to be $\lesssim0.1\%$. Observe that microreversibility demands comparing columns of (b) with rows of (c). The spectral decomposition and transition probabilities are calculated for the same parameters stated in Fig. 1.}
    \label{fig3}
\end{figure*}

\section{Forward and backward protocols}

Our specific proposal to test Eq. \ref{MR} is outlined in Figs. 2a-d. It involves the application of two separate protocols to the CPB to measure $P_{m|n} [\lambda]$ and $P_{n|m} [\tilde{\lambda}]$, which we refer to as the forward ($\lambda$) protocol and the backward ($\tilde{\lambda}$) protocol respectively. For process $\lambda$, the CPB is taken through the following sequence of steps: (1) First, with the external physical parameters set such that the energy eigenstates are definite charge states (i.e. $\beta\ll1$), the CPB is initialized in its ground state and driven by a pulse sequence consisting of the simultaneous application of time-varying signals $n_g (t)$ and $\Phi(t)$ (Figs. 2c-d), causing it to repeatedly pass through avoided-level crossings in its energy spectrum (Fig. 2e). At each such crossing, the CPB can undergo a Landau-Zener transition\cite{Zener,Nori} between the adjacent states involved in the crossing, which leaves it in a superposition of those two states. For the parameters considered here, after traversing the multiple crossings shown in Fig. 2e, the CPB state should be in a superposition of as many as 5 charge states. It should be noted that at the end of the sweep, $n_g$ and $\Phi$ are brought back to their initial values so that once again $\beta\ll1$. (2) At this point, immediately after the initial superposition state is prepared, a projective measurement of the CPB's charge is made and recorded as state $n$. We refer to those steps as the Preparation Protocol (Figs. 2a-b), since they provide an effective way for preparing an initial ensemble of charge states which can be used for measuring $P_{m|n}$ -in our case, its composition is given in (Fig. 3a). (3) Next, after the collapse to the charge state $\ket{n}$, a second pulse sequence identical to the sequence in step (1) is applied, again preparing the CPB in a superposition of charge states. (4) Finally, a second projective measurement of the CPB charge state is made and recorded as $m$. After step (4), the CPB is allowed sufficient time to relax back to its ground state, after which time $\lambda$ is repeated. In this manner, repeating $\lambda$ many times, the transitions probabilities $P_{m|n} [\lambda]$ can be constructed. Figure 3b illustrates a histogram of $P_{m|n} [\lambda]$ for this process calculated with numerical simulations using Eq. \ref{Hamiltonian} and the pulse sequences in Figs. 2c-d (See Appendix B).

To implement the time-reversed process $\tilde{\lambda}$ and construct the corresponding transition probabilities $P_{n|m} [\tilde{\lambda}]$, the same general procedure as outlined in the previous paragraph is followed. However, it is necessary to change two physical quantities for the time-reversed process: First, the sign of the magnetic flux applied to the CPB must be reversed to account for the reversal of momentum of the magnetic field's source charges. Observe that such inversion leads to ${\cal E}_J (-\Phi)={\cal E}_J^* (\Phi)$. Then, since the time reversal operator is an antilinear operator (see Appendix A), the system Hamiltonian is left invariant when taking the time reversal transformation together with the magnetic field sign change. Second, one should also invert the sign of the appropriate canonical variable of the CPB during $\tilde{\lambda}$, which in this case turns out to be the effective phase difference $\varphi$ across the CPB's Josephson junctions. Even though we have already found that the Hamiltonian is left invariant under the joint action of the time reversal operator and the magnetic field inversion,  this step must be done in order to preserve the time-reversal invariance of the canonical charge-phase commutation relations, since charge is considered an invariant under TRS. Furthermore, not performing such a transformation makes the time reversal transformation of the supercurrent density ill defined (see Appendix A). In our particular case, inverting the sign of $\varphi$ together with the antilinear transformation due to the time reversal operation has the effect of conjugating ${\cal E}_J$ in the Hamiltonian Eq. \ref{Hamiltonian}. Therefore, together with as one would expect, applying those two changes leaves the system Hamiltonian invariant. In addition, since we have a time-dependent Hamiltonian (nonautonomous system)\cite{Campisi}, we also have to revert the forcing protocol applied to the system, i.e., $|\Phi(t)|\rightarrow |\Phi(-t)|$ and $n_g (t)\rightarrow n_g (-t)$. With these changes, numerical simulations of the backward protocol indeed predict that Eq. 1 should hold (Fig. 3c).

\section{Conditions for unitarity and the measurement protocol}
To claim a true test of TRS through verification of microreversibility (Eq. \ref{MR}), it is essential that the CPB's time evolution be predominantly unitary during the $\lambda$ and $\tilde{\lambda}$ protocols. This requires that the protocols be implemented on a time scale $\tau_p$ that is much faster than any environmental effects. By applying the methodology introduced by Burkard-Koch-DiVincenzo\cite{BKD}, one finds that the figure of merit for quantifying such effects in our protocol is the relaxation time $T_1$. From those estimations, it can be shown that the decoherence time $T_2$ is determined by $T_1$ ($T_2\sim2T_1$), except for the regime $\beta\ll1$, which corresponds to a tiny window of $\sim0.2 {\rm ns}$ in the protocol, during which $T_2\sim 0.02 T_1$ (See Appendix C). Thus, even for a modest $T_1$ of 50ns, which is readily achievable with current technology\cite{Wallraff}, the designed protocol with $\tau_p\sim1 {\rm ns}$ (Figs. 2a-b) should provide a satisfactory unitary evolution.

It is also important that the projective charge measurements are made within a time-scale $\tau_{meas}\ll T_1$. That $T_1$ sets the relevant time-scale can be understood by realizing that decoherence effects becomes innocuous if one chooses projective measurements in the eigenenergy basis, since such effects would not lead to changes in the system state eigenenergy spectral decomposition. Notice from Fig. 2d that our protocol complies with this case: at the end of a protocol, when a projective measurement of charge is made, the CPB is biased so that the charge states are quasi-eigenenergy states of the system (i.e. $\beta\ll1$). Indeed, for the parameters chosen here (Fig. 1), each eigenenergy state has a probability larger than $99.8\%$ of being found in a specific charge state. Hence these measurements should also each be performed on a time scale $\tau_{meas}\lesssim10 {\rm ns}$. A natural and viable possibility for performing such high-speed, high-sensitivity charge measurements would be to use a superconducting single electron transistor (SSET)\cite{RFSET,Rimberg,Clerk}. When operated in RF mode, SSETs can have bandwidth in excess of 100 MHz \cite{RFSET} and charge detection sensitivity approaching the limit allowed by quantum mechanics \cite{Rimberg,Clerk}. Indeed, assuming the detection sensitivity achieved in Ref. \cite{Rimberg}, it should be possible to resolve the CPB's charge state with an error of $\sim0.5\%$ in a time scale of $t_{meas}\sim 20 {\rm ns}$ (See Appendix D). Such an error sets the precision limit for our proposal, since those due to the relaxation and dephasing processes impose a loss of state fidelity of the order of $1-\exp[-\int_0^{\tau_p}{dt/T_{(1,2)}(t)]}\sim-\int_0^{\tau_p}{dt/T_{(1,2)}(t)}$, for the short gate times under consideration. Using $T_1$ and $T_2$ time dependence determined\cite{BKD} for our protocol, one finds that the state fidelity degrades by $0.4\%$ (relaxation) and $0.3\%$(dephasing), by setting the conservative $T_1$ minimum value as $50{\rm ns}$.

\section{Gibbs ensemble emulation and quantum work fluctuation relation}
In addition to testing TRS in new macroscopic quantum limits, the investigations that we have outlined here would have implications for at least one contemporary avenue of investigation: quantum work fluctuation theorems\footnote{By work fluctuation theorems we mean relations between forward and backward probability distribution functions of physical quantities (e.g. work), for which the microreversibility principle is a necessary condition}. Indeed, to derive these theorems, it is necessary to make two hypotheses: the microreversibility principle and the assumption that the system is initially in thermal equilibrium at temperature $T$ (a Gibbsian distribution)\cite{Campisi}.

One paradigm of such work fluctuation theorems is the quantum Bochkov-Kuzovlev fluctuation theorem between the forward and backward work probability distribution functions (PDF)\cite{Campisi-BK}
\begin{equation}
\frac{P[W;\lambda]}{P[-W;\tilde{\lambda}]}=e^{W/k_BT}.
\label{PDF}
\end{equation}
Such a relation states that, when leaving an initial thermal equilibrium state, the system dynamics features a probability bias in favor of events for which work is done on the system  ($W>0$). Thus Eq. (\ref{PDF}) can been seen as a manifestation of the second law of thermodynamics, since it shows that energy releasing events are exponentially suppressed compared to energy absorbing events. The relation (\ref{PDF}) resembles very much the Tasaki-Crooks theorem\cite{Campisi}, for which the bias factor is $\exp[(W-\Delta F)/k_BT]$, where $\Delta F$ is the free energy difference between equilibrium thermal states associated with the system initial and final conditions. Those theorems are derived assuming different definitions of work, where one (Bochkov-Kuzovlev) associates work with the change in energy of the system unforced Hamiltonian, and the other relates it to changes in energy of the total system Hamiltonian (See Appendix E). Observe that both theorems give the same result for cyclical processes such as the protocol that we have proposed in this work.  It is important to stress that fluctuation theorems like Eq. (\ref{PDF}) are capable of determining the relative frequency with each of such events happen, which is a level of detail not provided by the standard thermodynamics approach of obtaining information from ensemble averages. Such a feature has been explored to understand and try to design quantum thermal machines\cite{Verley}. 

An immediate consequence of Eq. (\ref{PDF}) is the Bochkov-Kuzovlev equality $\langle e^{-W/k_B T}\rangle=1$\cite{Campisi-BK}, which clearly shows the power and the generality of the results derived from fluctuation theorems: independently of the specifications of the driving protocol and the characteristics of the system, the work distribution of any driving protocol applied to any system initially in thermal equilibrium at temperature $T$ is a random distribution, having the same expected value for the functional $\exp(-W/k_B T)$. 

To investigate quantum fluctuation theorems, like Eq. \ref{PDF}, using an SQD would require running an experiment at very low temperature ($T\sim30{\rm mK}$), in which case the SQD's initial thermal state is predominantly the population of the system ground state. Unfortunately this leads to very poor statistics for Eq. \ref{PDF}. In principle, this problem could be solved by just increasing the system temperature, but in order to obtain a Gibbsian distribution comprised of a reasonable number of states, e.g., 5 states, one should perform the experiment at $T\sim1{\rm K}$, at which temperature the SQD could no longer be well-approximated as a non-dissipative quantum system undergoing purely unitary evolution. 

\begin{figure}[t]
\begin{center}\includegraphics[ width=1\columnwidth,keepaspectratio]{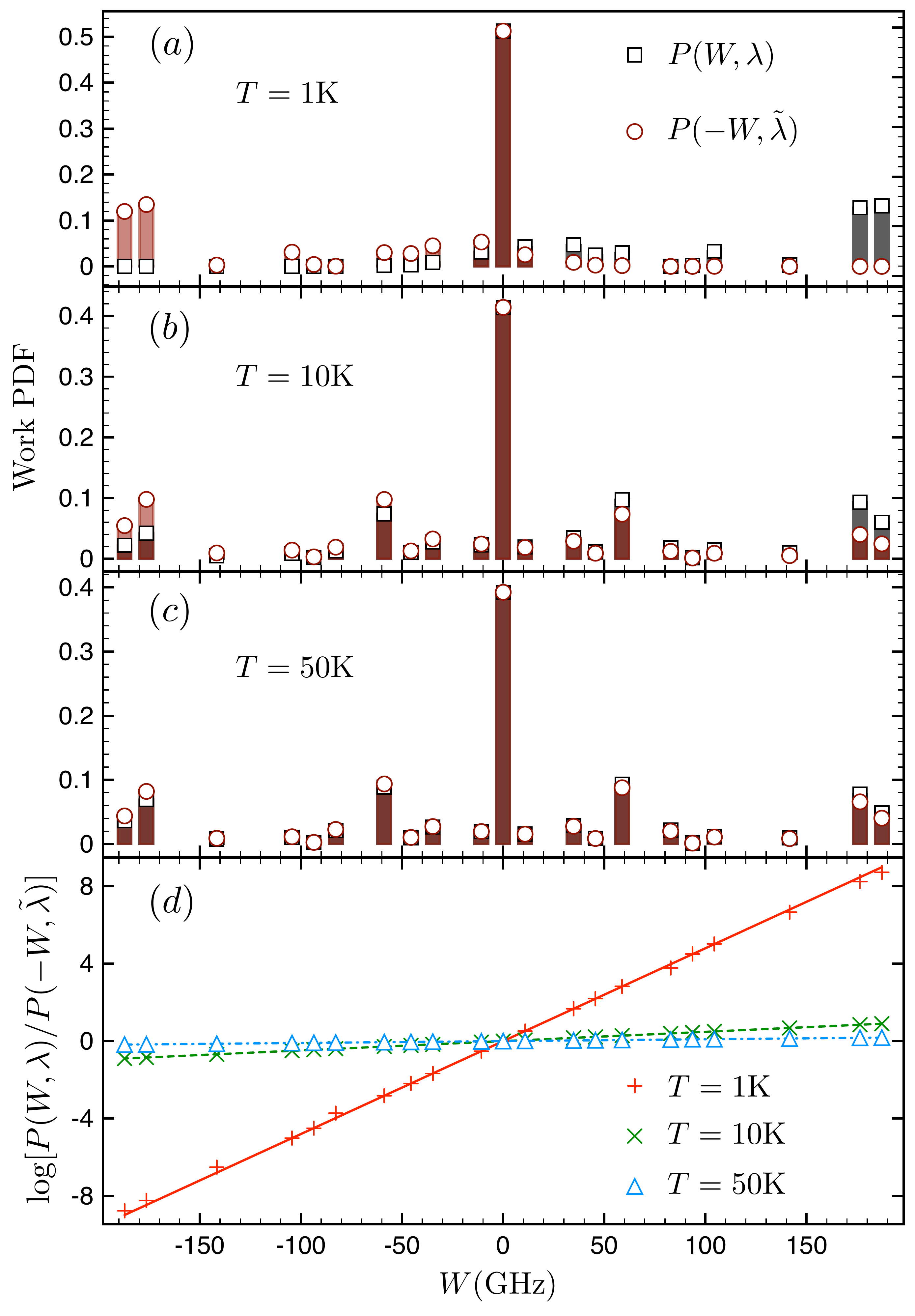}\end{center}
    \caption{Work probability distribution functions for the forcing protocol (Fig. \ref{fig2}), obtained from emulated Gibbs ensembles comprised of 5 states and generated from $10^6$ events. (a-c), Work PDFs for the forward $(P(W,\lambda))$ and backward $(P(-W,\tilde{\lambda}))$ protocols assuming temperatures of $T=1\rm K, 10\rm K,\rm{and} ~50\rm K$.  As depicted in panels (a-c), the higher the temperature, the closer become the forward and backward work PDFs. (d), The logarithmic plot of Eq. (\ref{PDF}) obtained from emulated Gibbs ensembles for different temperature $T$ values. Dot symbols are the obtained work PDF ratios and straight lines are guides to the eye representing the right hand side of Eq. (\ref{PDF}), which slopes are determined by the inverse of temperature.}
    \label{fig4}
\end{figure}

Here, we envision a solution to this problem by constructing a thermal state out of the initial charge ensemble obtained in the preparation protocol, which we name an emulation of an initial Gibbs ensemble. The procedure consists of randomly selecting the outcomes of the first measurement following the probability rule imposed by the Boltzmann weight $\exp(-H/k_B T)$. If the number of experimental events is sufficiently large, such distribution can be obtained for a given temperature $T$. Indeed, as Fig. \ref{fig4} and Table \ref{tab} show, for $N=10^6$ events, one can emulate a Gibbs ensemble comprised of those 5 states for temperatures above $1{\rm K}$, leading to $21$ possible different values of work, and verify with a small statistical error due to the sampling, which scales as $\sim1/\sqrt{N}$, the quantum Bochkov-Kuzovlev theorem and equality (See Appendix E).

Considering that, as presented to date, the system evolution in a quantum fluctuation theorem verification has to be disconnected from its environment during the forcing protocol\footnote{As a matter of fact, such a condition can be relaxed if one considers evolutions due to unital quantum channels, see: Alexey E. Rastegin, Non-equilibrium equalities with unital quantum channels. {\it J. Stat. Mech.} ({\bf 2013}) P06016.}, if one is not capable of monitoring the environment's state, the role played by the initial Gibbs ensemble in a such experimental verification is just to provide a set of initial states with their frequencies of appearance weighted by the Boltzmann factor - no system-environment correlation is maintained during single runs and between different runs of the experiment.  Thus, the ensemble emulation can be viewed playing the same role as a truly initial Gibbs ensemble in a quantum fluctuation theorem verification: a simple provider of uncorrelated initial states, the frequency of which is weighted by a know factor. Moreover, the emulation program can provide means to explore SQDs as quantum thermal machines\cite{Verley,PekolaNAT}.
\begin{table}[h]
\caption{Emulation for $10^6$ Events} 
\centering
\begin{tabular}{c rrrrrrr}
\hline\hline
Temperature $({\rm K})$&$1-\langle e^{-W/k_BT}\rangle$ \\ [0.5ex]
\hline 
1&$(-0.4\pm5.8)\times10^{-2}$\\ %
10 &$(-2.7\pm7.9)\times10^{-4}$ \\
20 &$(-2.1\pm4.2)\times10^{-4}$ \\
30 &$(-0.6\pm3.0)\times10^{-4}$ \\
40 &$(-0.1\pm2.2)\times10^{-4}$ \\
50 &$(-0.3\pm1.7)\times10^{-4}$ \\[1ex]
\hline
\end{tabular}
\label{tab}
\end{table}

\section{Conclusions}

In the present paper we have created a protocol for the preparation, time evolution and measurement of the quantum state of an SQD in order to test TRS in a new regime, namely in macroscopic quantum dynamics, using current technology and techniques. Our numerical simulations show that the repeated application of this protocol to the SQD would enable verification of the microreversibility principle in an artificial atom. 

Aside from being of fundamental importance to both equilibrium and non-equilibrium statistical mechanics, such a result would have the immediate consequence of verifying quantum fluctuation theorems via the construction of work probability distribution functions. This has been the subject of intense interest since the first proposals for determining the work PDF in quantum systems were put forth\cite{Huber,Heyl}. Recently the first experimental verifications have been accomplished in nuclear magnetic resonance\cite{Serra} and ion trap\cite{An} systems, but yet remaining an outstanding task for artificial atoms. In addition, several works have put forward the idea of inferring the work PDF using approaches that eliminate either the need of implementing successive projective energy measurements or the requirement of having an environment-isolated system dynamics - such as using Ramsey interferometry performed on an ancilla system\cite{Dorner, Mazzola, Hanggi}, single projective measurements of observables\cite{Fusco}, quantum jump measurements of a system and its environment in open quantum systems\cite{Hekking,Suomela}, or implementing the Positive Operator Valued Measure (POVM) technique \cite{Paz}. Notwithstanding that those approaches can represent great improvement for determining the work PDF in many systems, when considering the dynamics of macroscopic quantum states, the demand for the capability of either i) maintaining an auxiliary quantum system coherently coupled to the system of interest during the measurement protocol\cite{Dorner, Mazzola, Hanggi,Paz}, ii) restricting the investigation to sudden quench processes of specific initial states\cite{Fusco}, or iii) monitoring the environment's state in order to analyze the system-environment energy exchange\cite{Hekking}, may constitute requirements as difficult as the original task of performing successive projective energy measurements, in which our proposal is based on.

Although one could argue that our results are formally expected, their experimental observation would be of utmost importance and, apart from what we have said above, the reason is threefold. Firstly, it would provide the first direct test of the microreversibility of transitions between the states of a macroscopic quantum system. Secondly, it could further our understanding of how collective variables couple to their environment and lead to new techniques to enhance the reliability of decoupling from environmental degrees of freedom. Finally, if microreversibility is indeed observed in this kind of system it may constitute additional possible evidence of the applicability of quantum mechanics beyond its original realm.

\section{Acknowledgements}
FB and AOC are supported by Instituto Nacional de Ci\^encia e Tecnologia - Informa\c{c}\~ao Qu\^antica (INCT-IQ) and by Funda\c{c}\~ao de Amparo \`a Pesquisa do Estado de S\~ao Paulo  (FAPESP) under grant number 2012/51589-1. MDL and FR acknowledge support for this work provided by the National Science Foundation under Grant No. DMR-1056423 and Grant No. DMR-1312421.

\section{Appendix A: Time reversal symmetry in driven superconducting devices}
\subsection{The time reversal operation}
By definition, the effect of the time reversal operator $\Theta$ in mechanical systems is to reverse the linear $(\mathbf P)$ and orbital angular $(\mathbf L)$ momenta while leaving the position $\mathbf X$ unchanged, i.e., $\Theta\mathbf P\Theta^{-1}=-\mathbf P,\Theta\mathbf L\Theta^{-1}=-\mathbf L, ~{\rm and}~\Theta\mathbf X\Theta^{-1}=\mathbf X.$
For consistency, in order to extend the notion of time reversal for systems with spin variables, the spin angular momentum $\mathbf S$ must transform like the orbital angular momentum, i.e. $\Theta\mathbf S\Theta^{-1}=-\mathbf S$.

As for the electromagnetic phenomena, it is well known that the Maxwell equations and the Lorentz force are invariant under time reversal \cite{Jackson}. By choosing the convention that the electric charge is an {\it invariant} under time reversal, the TRS arises provided that the electric $\mathbf E$ and magnetic $\mathbf B$ field transformations are given by
\begin{eqnarray*}
\mathbf E\rightarrow \mathbf E~~{\rm and}~~\mathbf B\rightarrow -\mathbf B.
\end{eqnarray*} In addition, the current density $\mathbf j$ must reverse sign, i.e., $\mathbf j\rightarrow -\mathbf j$, which also conforms to its definition in terms of charge times velocity.

A key feature of the time reversal operator $\Theta$ is to be an antilinear operator. Such a property can be verified by inspection of the transformation of the canonical commutator, which reverses sign under TRS
\begin{equation*}
\Theta[X_\alpha,P_\beta]\Theta^{-1}=-[X_\alpha,P_\beta]=\Theta i\hbar\delta_{\alpha,\beta}\Theta^{-1}.
\label{commutator}
\end{equation*} 
Therefore it is necessary of $\Theta$ to be an antilinear operator, i.e. $\Theta i\Theta^{-1}=-i$, in order to preserve the commutation relations \footnote{That is also true when considering the commutation relations between the components of angular momentum, or between components of $\mathbf X$ and $\mathbf L$ or $\mathbf P$ and $\mathbf L$.}.

\subsection{Superconducting devices under time reversal}

The standard approach to quantize the dynamics of a superconducting circuit consists in elevating flux and charge variables to the status of operators \cite{Makhlin,Clarke}. Indeed, it can be shown that the superconducting phase difference $\varphi$ across a Josephson junction and the charge $Q$ on the junction capacitance are canonically conjungate variables \cite{BKD}. Therefore, it follows from the canonical quantization    that the conjugated variables $\varphi$ and $Q$ should obey commutation relations as
\begin{eqnarray*}
\left[\frac{\Phi_0}{2\pi}\varphi,Q\right]=i\hbar.
\end{eqnarray*} Since fundamental commutation relations should be preserved under time reversal, it must be determined which conjugated variable must change sign under time reversal, once that $\Theta i\hbar\Theta^{-1}=-i\hbar$. If one follows the standard approach of considering charge as invariant under time reversal\cite{Jackson}, then the transformation
\begin{eqnarray}
\Theta \varphi\Theta^{-1}=-\varphi~~{\rm and}~~ \Theta Q\Theta^{-1}=Q,
\label{Transformation}
\end{eqnarray} complies with the requirements. It should be appreciated that the above transformation is consistent with the expectation regarding the transformation of currents. Indeed, the supercurrent density can be written as
\begin{eqnarray*}
\mathbf j_s=i\frac{e\hbar}{2m_{\rm e}}\left\{\psi\nabla\psi^\ast-\psi^\ast\nabla\psi\right\}-\frac{2e^2}{m_{\rm e}}|\psi|^2\mathbf A,
\end{eqnarray*} where $2e$ and $2m_{\rm e}$ are respectively the charge and mass of a Cooper pair of electrons, $\mathbf A$ is the vector potential of any magnetic field applied, and $\psi$ represents the wave function of the macroscopic state occupied by the Cooper pair condensate  \cite{Tilley}. Then, if one writes $\psi({\mathbf r},t)=|\psi({\mathbf r},t)|\exp[i\phi({\mathbf r},t)]$, it is found that
 \begin{eqnarray*}
\mathbf j_s=\frac{e\hbar}{m_{\rm e}}|\psi|^2\nabla\phi-\frac{2e^2}{m_{\rm e}}|\psi|^2\mathbf A.
\end{eqnarray*} Consequently, if charge is taken invariant and the vector potential $\mathbf A$ reverses sign, $\mathbf j_s$ will {\it only} conform with the expectation of reversing sign under time reversal if the sign of the supercurrent phase $\phi$ is changed.

The time reversal of Hamiltonian (\ref{Hamiltonian}) is obtained according to the transformation rule Eq. (\ref{Transformation}). Since the charge state $\ket{n}$ is an eigenstate of $Q$ with real eigenvalue $n$, the invariance of $Q$ under time reversal implies that $Q(\Theta\ket{n})=\Theta Q\ket{n}=n(\Theta\ket{n})$. Noticing that $Q$ has nondegenerate eigenstates, it follows that $\Theta\ket{n}$ and $\ket{n}$ represent the {\it same} charge state and hence can differ at most by a constant phase, which can be set as $+1$ without loss of generality. The action of the antiunitary operator $\Theta$ on the charging energy and Josephson coupling leads  respectively to: $\Theta E_C(n-n_g)^2=E_C(n-n_g)^2\Theta$ and $\Theta{\cal E}_J(\Phi)={\cal E}_J^\ast(\Phi)\Theta$ \footnote{In order to directly verify the effect of the time reversal operation in $\varphi$ for Hamiltonian Eq. \ref{Hamiltonian}, it is instructive looking at the Josephson interaction Hamiltonian without choosing a specific representation, which reads $H_J(\Phi)=-({\cal E}_J(\Phi)e^{i\varphi}+{\cal E}_J^\ast(\Phi)e^{-i\varphi})/2$ and hence $\Theta H_J(\Phi)\Theta^{-1}=-({\cal E}_J^\ast(\Phi)e^{-i(-\varphi)}+{\cal E}_J(\Phi)e^{i(-\varphi)})/2$. Therefore time reversing $\varphi$ together with the antilinear transformation due to the time reversal operation has the effect of conjugating ${\cal E}_J$ in Eq. \ref{Hamiltonian}.} Thus, without inverting the sign of the applied magnetic field, one reaches the transformation\begin{multline}
H\rightarrow\Theta H\Theta^{-1}=4E_C\sum_n(n-n_g)^2\ket{n}\bra{n}\\
-\sum_n\left[\frac{{\cal E}_J^\ast(\Phi)}{2}\ket{n}\bra{n+1}+\frac{{\cal E}_J(\Phi)}{2}\ket{n+1}\bra{n}\right],
\end{multline} which only restores the original Hamiltonian when one reverses the applied magnetic field, since under this operation ${\cal E}_J^\ast(\Phi)\rightarrow{\cal E}_J^\ast(-\Phi)={\cal E}_J(\Phi)$.

\subsection{Time reversal symmetry of driven systems}

Despite the standard presentation of TRS as a feature of nondriven (autonomous) systems\cite{Messiah}, the concept of time-reversal invariance and the principle of microreversibility can be discussed in more general cases, where the system dynamics is driven by a time-dependent force. As we show below, the inversion of the Hamiltonian's temporal sequence is of prime feature when discussing TRS in driven systems (see \cite{Campisi} for a more detailed presentation).

The time reversal transformation of the Schr\"odinger equation yields\begin{gather*}
\Theta H(t)\Theta^{-1}\Theta\ket{\psi(t)}=\Theta\left(i\hbar\frac{\partial}{\partial t}\ket{\psi(t)}\right)=-i\hbar\frac{\partial }{\partial t}\Theta\ket{\psi(t)},\\
\Rightarrow H_{\rm{rev}}(t)\ket{\psi(t)}_{{\rm rev}}=i\hbar\frac{\partial }{\partial t}\ket{\psi(t)}_{{\rm rev}},
\end{gather*}
with $H_{\rm{rev}}(t)\equiv\Theta H(\tau-t)\Theta^{-1}$ and $\ket{\psi(t)}_{{\rm rev}}\equiv \Theta\ket{\psi(\tau-t)}$.  Observe that $H_{\rm{rev}}(t)$ and $\ket{\psi(t)}_{{\rm rev}}$ represent the system motion-reversed Hamiltonian and state, respectively. 

It is clear, then, that if $H$ is invariant under time reversal, i.e., $[\Theta,H(t)]=0,~\forall t$, the time evolution of the motion-reversed state $\ket{\psi(t)}_{{\rm rev}}$ is determined by the time-reversed image of $H(t)$, satisfying initial condition related to the state $\ket{\psi}$, namely \begin{eqnarray*}
\ket{\psi(0)}_{{\rm rev}}= \Theta\ket{\psi(\tau)}.
\end{eqnarray*} 
A system is said to be invariant under time reversal symmetry if $[\Theta,H(t)]=0,~\forall t$. For a such system, the time evolution operator and its motion-reversed are related through a simple identity\cite{Campisi}, namely, $U^\dagger(t,0)=\Theta^{-1}U_{\rm{rev}}(\tau,\tau-t)\Theta$, which allows one to derive the microreversibility principle for driven systems: \begin{multline*}
|\bra{m}U(\tau,0)\ket{n}|=|\bra{\tilde{n}}\Theta U^\dagger(\tau,0)\Theta^{-1}\ket{\tilde{m}}|=\\
=|\bra{\tilde{n}}U_{\rm{rev}}(\tau,0)\ket{\tilde{m}}|, ~{\rm with} \ket{\tilde{\alpha}}\equiv \Theta\ket{\alpha},
\end{multline*} for all $\ket{n}~\rm{and}~\ket{m}$. Equation \ref{MR} represents a short notation of the above stated microreversibility principle, where $\lambda$ and $\tilde{\lambda}$ are used to represent the system forward-in-time Hamiltonian's temporal sequence and its motion-reversed transformation, respectively.  

\section{Appendix B:  Numerical simulations}
The system's state time evolution was determined through numerical simulations of the unitary time-ordered  evolution operator due to the Hamiltonian Eq. \ref{Hamiltonian}. The calculation was performed taking into account an $N = 51$ charge dimensional Hilbert space. Considering the time discretization procedure and the Hilbert space truncation, we estimated a maximum relative error of $\sim0.05\%$ for the probabilities quoted in the main text. The specific flux and charge pulses used in our protocol read: $\Phi(t)=(\Phi_0/2)\cos(2\pi\times\frac{3}{2}\times t)$ and $n_g(t)=0.05-2\cos(2\pi\times\frac{3}{2}\times t)$, with time in unit of nanoseconds. \\

\section{Appendix C: Decoherence and relaxation rates for the CPB}
The methodology introduced by Burkard-Koch-DiVincenzo\cite{BKD} allows one to use circuit theory for describing the dissipative elements of the circuit with a bath of oscillators model, from which it is possible to estimate the dissipative effects for multilevel superconducting devices. From this modelling, the system-bath coupling derived is a functional of the charge number operator $n$. Therefore, it will only connect the system energy eigenstates that have at least one charge state in common in their spectral decomposition.  For the physical CPB parameters and the flux and charge protocols considered in our proposal, we found that only neighbouring eigenstates share one charge state in their spectral decomposition. Thus, with a good approximation, the dissipative process can be viewed as a sequence of multiple processes involving only two eigenstates. Under this perspective, one can obtain the relaxation ($T_1$) and decoherence ($T_2$) times concerning those two levels. In the Born-Markov approximation, the relation between $T_1$ and the pure dephasing $T_\phi$ is found to be 

{\footnotesize 
\begin{equation*}  
\frac{T_\phi}{T_1}\sim\frac{4|\bra{e_{k}}n\ket{e_{k+1}}|^2}{|\bra{e_{k}}n\ket{e_{k}}-\bra{e_{k+1}}n\ket{e_{k+1}}|^2}\frac{e_{k+1}-e_{k}}{2k_BT}\coth{\frac{e_{k+1}-e_{k}}{2k_BT}},
\end{equation*}}

{\noindent where $e_{k}$ is the instantaneous value of the eigenenergy state $k$, and $T_2^{-1}=T_1^{-1}/2+T_\phi^{-1}$. Performing the calculation of the matrix elements above for each time instant of our protocol, we found that the decoherence time $T_2$ is determined by $T_1$, i.e., $T_\phi\gg T_1$, except for the regime $\beta\ll1$, during which $T_2\sim 0.02 T_1$. Observe that for $\beta\ll1$ (the SQD charge regime), the charge operator $n$ almost commutes with the system Hamiltonian, which explains why $T_1$ becomes the longest time scale here.}

\section{Appendix D: Estimate of the CPB measurement uncertainty}
To estimate the CPB charge-state measurement uncertainty, we first assume that the CPB is probed using an SSET that is coupled to the CPB through a capacitance $C_C$.  It is further assumed that the SSET charge sensitivity $S_Q$ is dominated by the noise of the pre-amplifier used to read-out the SSET. In this case, for each Cooper-pair number state $N$, the inferred charge $(Q_C)$ on $C_C$ will have a Gaussian distribution $p_N(Q_C)$  with R.M.S. of $\sigma_Q=\sqrt{(S_Q/\tau_{meas} )}$, where $\tau_{meas}$ is the measurement time. We then define the measurement uncertainty through the use of the Kolmogorov (trace) distance \cite{Chuang}, which is given by $D(p_N (Q_C),p_{(N+1)} (Q_C))={(1/2)} \int{|p_N (Q_C)-p_{(N+1)} (Q_C)|dQ_C }$. The probability to correctly identify from which of two adjacent probability distributions $p_N (Q_C)$ and $p_{(N+1)} (Q_C)$ an outcome of a measurement $Q_C$  comes is thus given by $P_D=(1+D)/2$.  For example, using ${S_Q}^{{1/2}}=1.7 \mu e/\sqrt{{\rm Hz}}$, which was achieved in \cite{Rimberg}, a measurement time $\tau_{meas}=20 ns$, and realistic parameters for the total capacitance of the CPB island $C_{\Sigma}=6.5 {\rm fF}$(corresponding to ${{E_C}/{\hbar}} = 2\pi \times 3 {\rm GHz})$ and  $C_C=0.20 {\rm fF}$, we find $P_D=99.5 \%$, corresponding to a measurement uncertainty of $0.5 \%$. It is assumed that the SET measurement is pulsed off during the forward and backward protocols so that it does not serve as a strong source of dephasing.

\section{Appendix E: Quantum work and the Gibbs ensemble generator}
The quantum Bochkov-Kuzovlev theorem Eq. \ref{PDF} is derived\cite{Campisi-BK} considering the {\it exclusive viewpoint} for the definition of quantum work. Such definition considers that the quantum work performed in a specific process $\lambda$ is determined as the difference of the outcomes of the eigenenergy measurements of the unperturbed system Hamiltonian done at the initial and final process times. In our case, the Hamiltonian Eq. \ref{Hamiltonian} can be viewed as $H(t)=H_0+H_p[\lambda(t)]$, where the unperturbed Hamiltonian $H_0$ is set as $H(t=0)$, and the force-dependent Hamiltonian perturbation $H_p[\lambda(t)]$ is given by $H(t)-H_0$. As for the Tasaki-Crooks theorem, the {\it inclusive viewpoint} for the definition of work is adopted, which considers the outcomes of eigenenergy measurements of the total system Hamiltonian at the initial and final process times. 

The Gibbs ensemble emulation is constructed using a standard pseudorandom routine to select the states out of the initial ensemble obtained from the preparation protocol. The pseudorandom choice is weighted by the Boltzmann weight for a given temperature $T$.\\


\begin{thebibliography}{4}%
\makeatletter
\providecommand \@ifxundefined [1]{%
 \@ifx{#1\undefined}
}%
\providecommand \@ifnum [1]{%
 \ifnum #1\expandafter \@firstoftwo
 \else \expandafter \@secondoftwo
 \fi
}%
\providecommand \@ifx [1]{%
 \ifx #1\expandafter \@firstoftwo
 \else \expandafter \@secondoftwo
 \fi
}%
\providecommand \natexlab [1]{#1}%
\providecommand \enquote  [1]{``#1''}%
\providecommand \bibnamefont  [1]{#1}%
\providecommand \bibfnamefont [1]{#1}%
\providecommand \citenamefont [1]{#1}%
\providecommand \href@noop [0]{\@secondoftwo}%
\providecommand \href [0]{\begingroup \@sanitize@url \@href}%
\providecommand \@href[1]{\@@startlink{#1}\@@href}%
\providecommand \@@href[1]{\endgroup#1\@@endlink}%
\providecommand \@sanitize@url [0]{\catcode `\\12\catcode `\$12\catcode
  `\&12\catcode `\#12\catcode `\^12\catcode `\_12\catcode `\%12\relax}%
\providecommand \@@startlink[1]{}%
\providecommand \@@endlink[0]{}%
\providecommand \url  [0]{\begingroup\@sanitize@url \@url }%
\providecommand \@url [1]{\endgroup\@href {#1}{\urlprefix }}%
\providecommand \urlprefix  [0]{URL }%
\providecommand \Eprint [0]{\href }%
\providecommand \doibase [0]{http://dx.doi.org/}%
\providecommand \selectlanguage [0]{\@gobble}%
\providecommand \bibinfo  [0]{\@secondoftwo}%
\providecommand \bibfield  [0]{\@secondoftwo}%
\providecommand \translation [1]{[#1]}%
\providecommand \BibitemOpen [0]{}%
\providecommand \bibitemStop [0]{}%
\providecommand \bibitemNoStop [0]{.\EOS\space}%
\providecommand \EOS [0]{\spacefactor3000\relax}%
\providecommand \BibitemShut  [1]{\csname bibitem#1\endcsname}%
\let\auto@bib@innerbib\@empty
\bibitem [{Note1()}]{Note1}%
  \BibitemOpen
  \bibinfo {note} {By work fluctuation theorems we mean relations between
  forward and backward probability distribution functions of physical
  quantities (e.g. work), for which the microreversibility principle is a
  necessary condition}\BibitemShut {NoStop}%
\bibitem [{Note2()}]{Note2}%
  \BibitemOpen
  \bibinfo {note} {As a matter of fact, such a condition can be relaxed if one
  considers evolutions due to unital quantum channels, see: Alexey E. Rastegin,
  Non-equilibrium equalities with unital quantum channels. {\protect \it J.
  Stat. Mech.} ({\protect \bf 2013}) P06016.}\BibitemShut {Stop}%
\bibitem [{Note3()}]{Note3}%
  \BibitemOpen
  \bibinfo {note} {That is also true when considering the commutation relations
  between the components of angular momentum, or between components of
  $\protect \mathbf X$ and $\protect \mathbf L$ or $\protect \mathbf P$ and
  $\protect \mathbf L$.}\BibitemShut {Stop}%
\bibitem [{Note4()}]{Note4}%
  \BibitemOpen
  \bibinfo {note} {In order to directly verify the effect of the time reversal
  operation in $\varphi $ for Hamiltonian Eq. \ref {Hamiltonian}, it is
  instructive looking at the Josephson interaction Hamiltonian without choosing
  a specific representation, which reads $H_J(\Phi )=-({\protect \cal E}_J(\Phi
  )e^{i\varphi }+{\protect \cal E}_J^\ast (\Phi )e^{-i\varphi })/2$ and hence
  $\Theta H_J(\Phi )\Theta ^{-1}=-({\protect \cal E}_J^\ast (\Phi
  )e^{-i(-\varphi )}+{\protect \cal E}_J(\Phi )e^{i(-\varphi )})/2$. Therefore
  time reversing $\varphi $ together with the antilinear transformation due to
  the time reversal operation has the effect of conjugating ${\protect \cal
  E}_J$ in Eq. \ref {Hamiltonian}.}\BibitemShut {Stop}%
\end{thebibliography}%


\begin{thebibliography}{99} 

\bibitem{Gross} Gross, D. J. The role of symmetry in fundamental physics. {\it Proc. Nat. Acad. Sci.} {\bf 93,} 14256 (1996).

\bibitem{Messiah} Messiah, A. {\it Quantum Mechanics} (Dover Publications, Mineola, 1999).

\bibitem{Reichl} Reichl, L. E. {\it A Modern Course In Statistical Physics}, second edition (John Wiley \& Sons, New York, 1998).

\bibitem{Callen} Callen, H. B. \& Welton, T. A. Irreversibility and generalized noise. {\it Phys. Rev. E} {\bf 83,} 34 (1951).

\bibitem{Campisi} Campisi, M., H\"anggi, P. \& Talkner, P. Quantum fluctuation relations: Foundations and applications. {\it Rev. Mod. Phys.} {\bf 83,} 771 (2011).

\bibitem{BABAR} Lees, J. P. {\it et al.} Observation of Time-Reversal Violation in the $B^0$ Meson System. {\it Phys. Rev. Lett.} {\bf 109,} 211801 (2012).

\bibitem{Bernabeu} Bernab\'eu, J. Time reversal violation for entangled neutral mesons. {\it Journal of Physics: Conference Series} {\bf 447,} 012005 (2013).

\bibitem{Sakharov} Sakharov, A. D. Violation of CP invariance, C asymmetry and baryon asymmetry of universe. {\it JETP Letters} {\bf 5,} 24 (1967).

\bibitem{henley} Henley, E.M. Time Reversal Symmetry. {\it Int. J. Mod. Phys. E} {\bf 22,} 1330010 (2013).

\bibitem{LeggettJP} Leggett, A. J. Testing the limits of quantum mechanics: motivation, state of play, prospects.
 {\it J. Phys.: Condens. Matter} {\bf 14,} R415 (2002).

\bibitem{Ball} Ball, P. {\it Nature} {\bf 453,} 22 (2008).

\bibitem{Clarke} Clarke, J. \& Wilhelm, F. K. Superconducting quantum bits. {\it Nature} {\bf 453,} 7198 (2008).

\bibitem{Devoret} Devoret, M.H. \& Shoelkopf, R.J. Superconducting circuits for quantum information: An outlook. {\it Science} {\bf 339,} 1169 (2013).

\bibitem{Cleland} O'Connell, A. D. {\it et al.} Quantum ground state and single-phonon control of a mechanical resonator. {\it Nature} {\bf 464,} 697 (2010).

\bibitem{Palomaki} Palomaki, T. A., Teufel, J. D., Simmonds, R. W. \& Lehnert, K. W. Entangling mechanical motion with microwave fields. {\it Science} {\bf 342,} 710 (2013).

\bibitem{Haroche} Raimond, J.M., Brune, M. \& Haroche, S. Colloquium: Manipulating quantum entanglement with atoms and photons in a cavity. {\it Rev. Mod. Phys.} {\bf 73,} 565 (2001).

\bibitem{Mooij} Mooij, J. E. {\it Nat. Phys.} {\bf 6,} 401 (2010).

\bibitem{Houck} Houck, A.A., T\"ureci, H.E. \& Koch, J. On-chip quantum simulation with superconducting circuits.{\it Nat. Phys.} {\bf 8,} 292 (2012).

\bibitem{Leggett} Leggett, A. J. Macroscopic Effect of {\it P}- and {\it T}-Nonconserving Interactions in Ferroelectrics: A Possible Experiment?. {\it Phys. Rev. Lett.} {\bf  41,} 586 (1978).

\bibitem{Ramsey} Ramsey, N. F. Electric-dipole moments of elementary particles. {\it Rep. Prog. Phys.} {\bf 45,} 95 (1982). 

\bibitem{Budker} Budker, D., Lamoreaux, S. K., Sushkov, A. O. \& Sushkov, O. P. Sensitivity of condensed-matter {\it P}- and {\it T}-violation experiments. {\it Phys. Rev. A} {\bf 73,} 022107 (2006).

\bibitem{Eckel} Eckel, S., Sushkov, A. O., \& Lamoreaux, S. K. Limit on the Electron Electric Dipole Moment Using Paramagnetic Ferroelectric Eu$_{0.5}$Ba$_{0.5}$TiO$_3$. {\it Phys. Rev. Lett.} {\bf 109,} 193003 (2012).

\bibitem{Makhlin} Makhlin, Y., Sch\"on, G. \& Shnirman, A. Quantum-state engineering with Josephson-junction devices. {\it Rev. Mod. Phys.} {\bf 73,} 357 (2001).

\bibitem{Blatter} Blatter, G. Schr\"odinger's cat is now fat. {\it Nature} {\bf 406,} 25 (2000).

\bibitem{CaldeiraBook} Caldeira, A. O. {\it An Introduction to Macroscopic Quantum Phenomena and Quantum Dissipation} (Cambridge University Press, Cambridge, 2014).

\bibitem{Sun} Sun, L. et al. Tracking photon jumps with repeated quantum non-demolition parity measurements. {\it Nature} {\bf 511,} 444 (2014).

\bibitem{Zener} Zener, C. Non-Adiabatic Crossing of Energy Levels. {\it Proc. R. Soc. Lond. A} {\bf 137,} 696 (1932).

\bibitem{Nori} Shevchenko, S.N., Ashhab, S., \& Nori, F. Landau-Zener-St\"uckelberg interferometry. {\it Phys. Rep.} {\bf 492,} 1 (2010).

\bibitem{BKD} Burkard, G., Koch, R. H. \& DiVincenzo, D. P. Multilevel quantum description of decoherence in superconducting qubits. {\it Phys. Rev. B} {\bf 69,} 064503 (2004).

\bibitem{Wallraff} Wallraff, A. et al. Strong coupling of a single photon to a superconducting qubit using circuit quantum electrodynamics. {\it Nature} {\bf 431,} 162 (2004).

\bibitem{RFSET} Schoelkopf, R.J. et al. The radio-frequency single-electron transistor(RF-SET): A fast and ultrasensitive electrometer. {\it Science} {\bf 280,} 1238 (1998).

\bibitem{Rimberg} Xue, W..W et al. Measurement of the quantum noise of a single-electron transistor near the quantum limit. {\it Nat. Phys.} {\bf 5,} 660 (2009).

\bibitem{Clerk} Clerk, A.A., Girvin, S.N., Ngyuen, A.K. \& Stone, A.D. Resonant Cooper-pair tunneling: Quantum noise and measurement characteristics. {\it Phys. Rev. Lett.} {\bf 89,} 176804 (2002).

\bibitem{Campisi-BK} Campisi, M., Talkner, P. \& H\"anggi, P. Quantum Bochkov-Kuzovlev work fluctuation theorems. {\it Phil. Trans. R. Soc. A} {\bf 369,} 291 (2011).

\bibitem{Verley} Verley, G., Esposito, M., Willaert, T. \& Van den Broeck, C. The unlikely Carnot efficiency. {\it Nat. Commun.} {\bf 5}, 5721 (2014).

\bibitem{PekolaNAT} Jukka P. Pekola. Towards quantum thermodynamics in electronic circuits. {\it Nat. Phys.} {\bf 11}, 118 (2015).

\bibitem{Huber} Huber, G. \& Schmidt-Kaler, F. Employing Trapped Cold Ions to Verify the Quantum Jarzynski Equality. {\it Phys. Rev. Lett.} {\bf 101}, 070403 (2008).

\bibitem{Heyl} Heyl, M. \& Kehrein, S. Crooks Relation in Optical Spectra: Universality in Work Distributions for Weak Local Quenches. {\it Phys. Rev. Lett.} {\bf 108}, 190601 (2012).

\bibitem{Serra} Batalhao, T. B. et al. Experimental reconstruction of work distribution and verification of fluctuation relations at the full quantum level. {\it Phys. Rev. Lett.} {\bf 113}, 140601 (2014).

\bibitem{An} An, S. et al. Experimental test of the quantum Jarzynski equality with a trapped-ion system. {\it Nat. Phys.} {\bf Advance online publication} (2014).

\bibitem{Dorner} Dorner, R. et al. Extracting Quantum Work Statistics and Fluctuation Theorems by Single-Qubit Interferometry. {\it Phys. Rev. Lett.} {\bf 110}, 230601 (2013).

\bibitem{Mazzola} Mazzola, L., De Chiara, G. \& Paternostro, M. Measuring the Characteristic Function of the Work Distribution. {\it Phys. Rev. Lett.} {\bf 110}, 230602 (2013).

\bibitem{Hanggi} Campisi, M. et al. Employing circuit QED to measure non-equilibrium work fluctuations. {\it New J. Phys.} {\bf 15}, 105028 (2013).

\bibitem{Fusco} Fusco, L. et al. Assessing the Nonequilibrium Thermodynamics in a Quenched Quantum Many-Body System via Single Projective Measurements. {\it Phys. Rev. X} {\bf 4}, 031029 (2014).

\bibitem{Suomela} Suomela, S. et al. Moments of work in the two-point measurement protocol for a driven open system. {\it Phys. Rev. B} {\bf 90}, 094304 (2014).

\bibitem{Hekking} Hekking, F. W. J. \& Pekola, J. P. Quantum Jump Approach for Work and Dissipation in a Two-Level System. {\it Phys. Rev. Lett.} {\bf 111}, 093602 (2013).

\bibitem{Paz} Roncaglia, A. J., Cerisola, F. \& Paz, J. P. Work measurement as a generalized quantum measurement. {\it Phys. Rev. Lett.} {\bf 113}, 250601 (2014).

\bibitem{Jackson} Jackson, J. D. {\it Classical Electrodynamics} (John Wiley \& Sons, New York, 1999).

\bibitem{Tilley} Tilley, D. R. \& Tilley, J. {\it Superfluidity and Superconductivity} (Institute of Physics Publishing, London, 2003).

\bibitem{Chuang} Nielsen, A. C. \& Chuang, I. L. {\it Quantum Computation and Quantum Information} (Cambridge University Press, Cambridge, 2000).

\end{thebibliography}
\end{document}